\documentclass[12pt]{article}
\usepackage{graphicx}
\usepackage{amsmath}

\newcommand\pubdate{\today}

\def\anl{Argonne National Laboratory, High Energy Physics Division\\
9700 S Cass Ave, IL 60439, USA}

\def\Title#1{\begin{center} {\Large #1 } \end{center}}
\def\Author#1{\begin{center}{ \sc #1} \end{center}}
\def\Address#1{\begin{center}{ \it #1} \end{center}}

\newcommand\pubblock{\rightline{\begin{tabular}{l} \\
         \pubdate  \end{tabular}}}
\newenvironment{Abstract}{\begin{quotation}  }{\end{quotation}}
\newenvironment{Presented}{\begin{quotation} \begin{center} 
             PRESENTED AT\end{center}\bigskip 
      \begin{center}\begin{large}}{\end{large}\end{center} \end{quotation}}
\def\Acknowledgements{\bigskip  \bigskip \begin{center} \begin{large}
             \bf ACKNOWLEDGEMENTS \end{large}\end{center}}

\def\beq{\begin{equation}}
\def\eeq#1{\label{#1}\end{equation}}
\def\eeqn{\end{equation}}

\def\beqa{\begin{eqnarray}}
\def\eeqa#1{\label{#1}\end{eqnarray}}
\def\eeqan{\end{eqnarray}}

\let\bar=\overbar

\def\Dslash{\not{\hbox{\kern-4pt $D$}}}
\def\dslash{\not{\hbox{\kern-2pt $\del$}}}

\def\msb{{\bar{\ssstyle M \kern -1pt S}}}

\begin{document}
\begin{titlepage}
\pubblock

\vfill
\Title{Experiences from the Commissioning and First Physics Run of the Fermilab Muon g-2 Experiment}
\vfill
\Author{ Ran Hong\\on behalf of the Muon g-2 Collaboration}
\Address{\anl}
\vfill
\begin{Abstract}
The Muon g-2 Experiment (E989) at Fermilab is seeking to measure the
anomalous magnetic moment of muon ($a_{\mu}$) with a precision of 140
parts-per-billion (ppb) and aiming to resolve the discrepancy between
the E821 measurement and the Standard Model calculation of
$a_{\mu}$. In E989, the muon beam is stored in a ring magnet. The anomalous spin
precession frequency $\omega_{a}$ is measured by counting decay
positrons in 24 calorimeters, and the magnetic field is measured by
nuclear magnetic resonance (NMR) probes. Improvements in this experiment
with respect to its predecessor and the progress achieved in the commissioning
run and the first physics run, Run-1, are presented.
\end{Abstract}
\vfill
\begin{Presented}
Thirteenth Conference on the Intersections of Particle and Nuclear Physics\\
Palm Springs, CA, USA,  May 29 -- June 3, 2018
\end{Presented}
\vfill
\end{titlepage}
\def\thefootnote{\fnsymbol{footnote}}
\setcounter{footnote}{0}

\section{Introduction}

\begin{figure}[htb]
\centering
\includegraphics[width=0.7\linewidth]{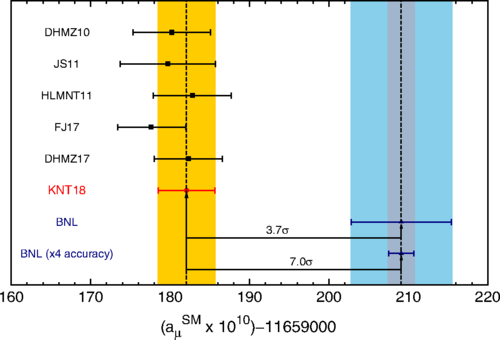}
\caption{Recent theoretical calculations of the anomalous magnetic moment of muon $a_{\mu}$ in the Standard Model.  The result of the experiment E821 and the projected uncertainty of E989 are also shown. This figure was originally presented in Reference~\cite{PhysRevD.97.114025}.}
\label{Fig_TheorySummary}
\end{figure}

The $g$-factor of a particle is the ratio of its magnetic moment (measured in $\mu_{B}\equiv q/2m$) and spin, and for spin-$1/2$ elementary particles $g=2$ if no quantum corrections are taken into account. Due to quantum corrections, the $g$-factor deviates from 2 \cite{PhysRev.73.416}, and the lepton magnetic anomaly $a=(g-2)/2$ describes the amount of the deviation. The part-per-billion level agreement between the measured electron anomaly $a_{e}$ and the prediction of it based on the Standard-Model (SM) physics has been a benchmark for the validity of quantum electrodynamics (QED) \cite{PhysRevLett.100.120801}. For muons, $a_{\mu}$ was measured (in experiment E821) with a precision of 540~ppb at Brookhaven National Lab \cite{PhysRevD.73.072003}. In 2006, the final result of E821 was different from the SM prediction by 2.5 standard deviations. This difference motivated the physics community to search for mechanisms beyond the SM \cite{PhysRevD.98.055015} that can explain the difference, as well as more accurate calculations of the contributions to $a_{\mu}$ from the known physics within the SM \cite{PhysRevD.97.114025}. Recently there have been great progresses in calculating the contributions from hadronic vacuum polarization (HVP) diagrams and hadronic light-by-light (HLbL) diagrams. The uncertainty on the HVP contribution was improved by using the dispersion-relation approach \cite{Jegerlehner:2017gek} with more recent experimental data \cite{DAVIER2014123}, while the HLbL contribution was better estimated using Lattice-QCD \cite{PhysRevD.96.034515}. Recently Lattice-QCD calculations for HVP and dispersion-relation approach for HLbL are also developed \cite{Meyer:2018til,Colangelo2017,PhysRevLett.121.112002}.  Recent results \cite{PhysRevD.97.114025} are shown in Figure~\ref{Fig_TheorySummary}, and the difference between the theoretical calculations and the previous measured value persists. 

The Muon g-2 Experiment at Fermilab (E989) \cite{Grange:2015fou} is
aiming at reducing the experimental uncertainty down to 140~ppb, a
factor of 4 improvement from E821. The new result will help resolve
the $a_{\mu}^{\text {SM}}-a_{\mu}^{\text {EXP}}$ discrepancy with
higher confidence. To achieve this goal, E989 will record 21 times the
E821 data set, while various sources of systematic uncertainties will be
reduced by improved instrumentation. In this paper, an overview of
the $a_{\mu}$ measurement principles are presented in
Section~\ref{Sec_Principle}. The E989 experiment design and
commissioning progress are described in
Section~\ref{Sec_Commission}. The current progress in Run-1 is
described in Section~\ref{Sec_PhysRun1}.

\section{Measurement principles}
\label{Sec_Principle}

In E989, a polarized muon beam is injected into a storage ring with a
uniform magnetic field. Because $g\neq2$ the spin precession angular
velocity $\vec{\omega}_s$ is different from the cyclotron angular
velocity $\vec{\omega}_c$. Assuming that the magnetic field is
perfectly uniform and betatron oscillations of the beam are neglected,
the difference of $\vec{\omega}_s$ and $\vec{\omega}_c$ is
\cite{PhysRevD.73.072003}:
\begin{align}
\vec{\omega}_a=\vec{\omega}_s-\vec{\omega}_c=-\frac{e}{m_{\mu}}\left[a_{\mu}\vec{B}-\left(a_{\mu}-\frac{1}{\gamma^2-1}\right)\frac{\vec{\beta}\times\vec{E}}{c}\right],
\label{Eq_SpinPrecession}
\end{align}
where $\vec{B}$ and $\vec{E}$ are the magnetic and electric fields
in the storage ring, $\vec{\beta}=\vec{v}/c$ is the velocity of
the muon relative to the speed of light, and
$\gamma=1/\sqrt{1-\beta^2}$. In the muon storage ring a static electric
field is used to focus the muon beam vertically. In order to reduce
the magnitude of the $\vec{\beta}\times\vec{E}$ term in
Eq.~\ref{Eq_SpinPrecession}, the momentum of the muon beam is chosen
to be 3.094~GeV/c ($\gamma=29.3$) so that the coefficient of this term
vanishes. In this experiment the {\em anomalous precession angular
  frequency} $\omega_{a}$ and the average magnetic field strength
$\tilde{B}$ experienced by the muons are measured separately, and the
anomalous magnetic moment is proportional to the ratio of them:
\begin{align}
a_{\mu}=-\frac{m_{\mu}\omega_{a}}{e\tilde{B}}.
\label{Eq_a_mu}
\end{align}

$\omega_{a}$ is extracted from the rate of positrons emitted in the
muon decay $\mu^{+}\rightarrow e^{+}\nu_{e}\bar{\nu}_{\mu}$. Due to
parity violation in weak interactions, the angular distribution of
the emitted positrons is not isotropic. In the rest frame of a
positive muon, high-energy positrons are more likely to be emitted
along the muon spin. As the spin of the muon precesses in the magnetic
field, the asymmetric angular distribution of the positrons also
rotates at the angular frequency $\omega_{a}$. Because the kinetic
energy of the muons in the lab frame is much higher than the rest-frame end-point
energy of the emitted positrons, the positrons experience a huge
Lorentz boost. Therefore, most of the positrons are emitted in the
forward direction, and their lab-frame energies $E_{lab}$ strongly
depend on their emission angle $\theta^{*}$ in the rest frame relative
to the boost direction
\begin{align}
E_{\text lab} \approx \gamma E^{*}(1+\cos \theta^{*})
\end{align}
where $E^{*}$ is the positron energy in the rest frame of the
muon. Through this mechanism, $E_{\text lab}$ also oscillates at the
angular frequency $\omega_{a}$. The emitted positrons have less
momenta than the muons, so their orbit radius in the magnet is smaller
than the muon storage orbit radius and they curve towards the inner
side of the ring and eventually hit the calorimeters that measure the
arrival time and energy deposition of the incoming particles. From the
calorimeter readings, one can reconstruct the energy deposition versus
time and extract its oscillation angular frequency (Q-method), or set
a threshold on energy and reconstruct the positron count versus time
and extract its oscillation angular frequency (T-method). These two
methods have different systematic uncertainties, and in E989 both of
them are implemented by different analysis teams, and their results
will be compared and evaluated before the final result is released.

The magnetic field in the muon storage region is mapped using proton
nuclear magnetic resonance (NMR) probes. The NMR system determines the
free-proton precession angular frequency $\omega_{p}$ which is
proportional to the magnitude of the magnetic field. The average
magnetic field $\tilde{\omega}_{p}$ experienced by the muons is
obtained by integrating the $\omega_{p}$ map weighted by the measured
muon distribution map. To express $a_{\mu}$ in terms of $\omega_{a}$
and $\tilde{\omega}_{p}$, Eq.~\ref{Eq_a_mu} becomes
\begin{align}
a_{\mu}=\frac{g_{e}}{2}\frac{m_{\mu}}{m_{e}}\frac{\mu_{p}}{\mu_{e}}\frac{\omega_{a}}{\tilde{\omega}_{p}},
\end{align}
where $g_{e}$ is the $g$-factor of an electron, $m_{\mu}/m_{e}$ is the
muon-to-electron mass ratio, and $\mu_{p}/\mu_{e}$ is the
proton-to-electron magnetic moment ratio, and these three values are
already measured with uncertainties better than 22~ppb in experiments
\cite{PhysRevLett.100.120801,PhysRevLett.82.711,RevModPhys.88.035009}.

\section{Overview of the E989 commissioning progress}
\label{Sec_Commission}

The construction of E989 started in 2013, and in May 2017 most of the
major components were built and installed. After 11 months of studying
and improving the experiment, E989 was commissioned and Run-1 started
at the end of March 2018. In this section, key components of the
experiment will be described, together with the improvements compared
to E821 and the milestones achieved during the commissioning period.

\subsection{Muon beam line}

The muon beam \cite{PhysRevAccelBeams.20.111003} used in E989 is
produced by the Fermilab accelerator complex. Hydrogen ions generated
at the Ion Source are accelerated to 400~MeV in the Linear Accelerator
and then boosted to 8~GeV in the Booster. The proton beam from the
Booster is then directed to the Recycler for bunching, and then
impinged on a Inconel target to create pions. Positive pions with 3.11
GeV/c momentum are selected and directed into the Delivery Ring which
was used as a part of the anti-proton source when the Tevatron was
in use. Polarized muons with 3.094~GeV/c momentum are captured from
forward-going pion decays. In the delivery ring, the beam goes for
several turns while more pions decay into muons, and the muon bunch is
separated spatially from the remaining protons so that a clean muon
bunch can be extracted. The extracted muon bunches are delivered to
the muon storage ring in the Muon g-2 Experiment Hall. In May 2017,
all the beamlines except the Delivery Ring were commissioned. The
latter was commissioned in December 2017. In the early runs from May
to December 2017 protons and pions were injected together with muons
into the storage ring, and since December 2017 protons and pions were
eliminated before injection. Through the commissioning run, the muon
beam was tuned and improved to optimize the number of muons delivered
to the experiment. In May 2018, 300,000 muons per bunch were being recorded
just prior to injection through the storage ring yoke, which is close to the goal set in the {\em Technical Design Report}  \cite{Grange:2015fou}.

\subsection{Muon storage ring}

The superconducting ring magnet provides a 1.45~T uniform magnetic
field to store the muons. It is the same magnet \cite{DANBY2001151}
used in E821, and it was transported to Fermilab in the summer of
2013. The superconducting coils with their cryostats were transported
as a whole, while the iron pieces were disassembled and transported
separately and then reassembled. The magnet was successfully cooled
and powered in 2015, and since then procedures for operating the
magnet safely and stably were developed.

\begin{figure}[htb]
\centering
\includegraphics[width=0.9\linewidth]{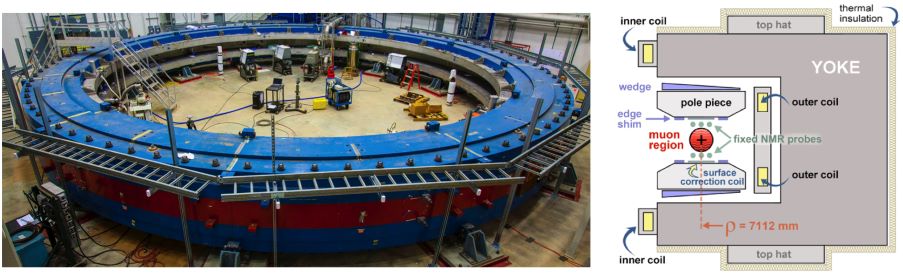}
\caption{The photo of the storage ring magnet and its cross-sectional
  view. The circular region between the poles represents the muon
  storage region. The magic radius of $\rho$=7112~mm relative to the
  center of the ring is labeled. The superconducting coils are marked
  by the rectangles at the inner and outer radii.}
\label{Fig_MagnetCrossSection}
\end{figure}

The cross-section of the storage ring magnet is shown in
Figure~\ref{Fig_MagnetCrossSection}. The magnetic field points
vertically upwards in the space between the pole pieces. To achieve a
better uncertainty on $\tilde{\omega}_p$, the magnet has to be shimmed
to a higher uniformity than that was the case in E821. In the magnet shimming
campaign of E989, each of the adjustable iron pieces (top/bottom hats,
pole pieces, edge shim and wedge shims) was adjusted carefully to
reduce the transverse (radial and vertical directions) and
longitudinal (azimuthal direction in the ring) gradients of the
field. To shim the non-uniformity at even shorter scales than the
sizes of these iron pieces, customized iron foils were used to increase
the field strength where the field was weak. $\sim$8500 precisely cut iron foils
were epoxied to the surface of the pole
pieces. It took 10~months to finish the shimming, and the peak-to-peak
variation of the field reached $\pm$25~ppm in the azimuthal direction
and $\pm$4~ppm in the transverse directions
\cite{Matthias2017}. Besides these passive shimming techniques, 200
concentric coils are placed on the surface of the pole pieces, and the
current in the coils are programmable in order to cancel the remaining
transverse field non-uniformity. After optimizing the current in the
coils, the peak-to-peak variation of the azimuthally averaged field
cross-sectional map was reduced to 2.5~ppm. The current in the main
magnet coils can be actively adjusted based on the field measurements
by the fixed probes. This feedback mechanism maintains the field at a
constant value over a long time period.

Twelve sections of the storage ring vacuum chamber inherited from E821
were cleaned and installed by March 2017. Some chambers were modified
to host new devices like the in-vacuum straw trackers. Inside the
vacuum chamber all around the ring, there are rails guiding the
magnetic field scanning device. Because the rails determine the radial
and vertical coordinates of the field scanner, the positions and
shapes of rails in each chamber section were aligned at the
sub-millimeter level.

The muon beam enters the magnet through an aperture in the back of the
iron yoke. From the inner edge of the iron yoke to the edge of the pole pieces, he magnetic field increases from near zero to the full field strength. To prevent the beam from being
deflected by the main field before reaching the storage region, a
superconducting magnet called the {\em inflector} is used to cancel
the main magnetic field along the path of the muon beam between the
cryostat of the outer magnet coils and the storage ring vacuum. It has
a superconducting shield so that the field generated by the inflector
currents does not affect the highly uniform field in the beam storage
region. The inflector used in E821 was installed and has been operated since
December 2016. A new inflector magnet with an open-ended design is
being built for FY20 running. The open-ended inflector does not
scatter muons at the exit and so will improve the muon storage
efficiency.

\begin{figure}[htb]
\centering
\includegraphics[width=0.8\linewidth]{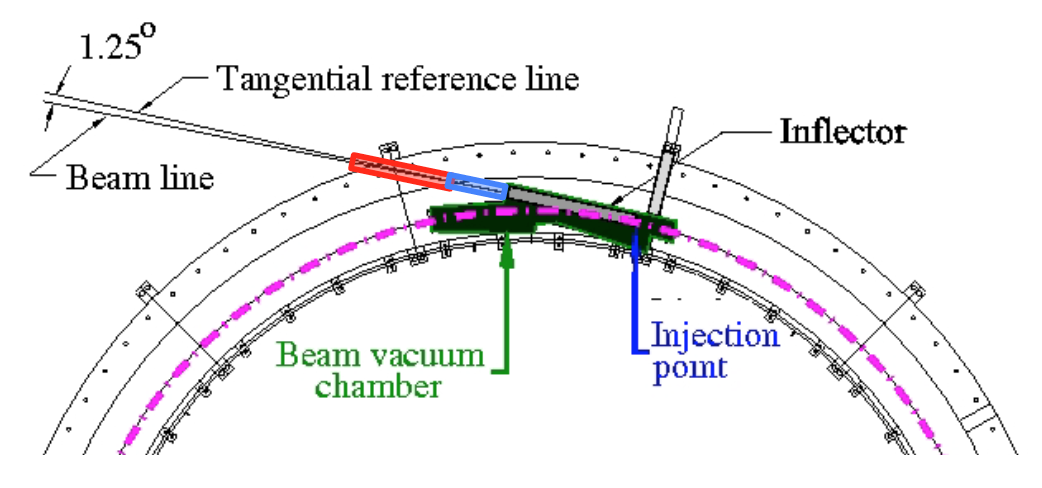}
\caption{Schematics of the muon beam entering the storage ring through
  the tunnel in the iron and then the inflector.}
\label{Fig_Inflector}
\end{figure}

When the muons exit the inflector they are displaced radially 77~mm outward from
the central (ideal) orbit of the storage ring, and therefore after
injection the muon orbit is a circle offset by the same amount from
the ideal orbit. The muons cross the ideal orbit approximately
90$^{\circ}$ azimuthally downstream from the end of the inflector. At
this point, the direction of the muons are not tangential to the ideal
orbit. The kicker magnet imposes a fast magnetic field pulse to the
incoming muon bunch, and bends them onto the ideal
orbit. There are three kickers in the E989 storage ring, and each of
them is a pair of metal plates on the inner and outer radius of
the beam. A $\sim$5000~A current is conducted through
the plates creating a transient vertical field peaked at $\sim$250
Gauss. The ideal pulse shape of the kicker is square with a flat top
that lasts for $\sim$100~ns. This ideal pulse was not achieved during the commissioning run and significant work was undertaken to improve the characteristics of the pulse and a larger kicker pulse will be available in the upcoming FY19 run.


Electrostatic quadrupoles are used in the storage ring to confine the
muon beam vertically, and they cover $\sim$43\% of the total
circumference. The electrodes from E821 were polished, aligned and
installed into the vacuum chambers early in 2017. The electrodes are
aluminum plates. When the muon beam moves from the inflector to the
kicker for the first time, it passes through the outer electrode.  In
E989 this electrode is made of aluminized mylar in order to reduce the
beam scattering and thus improve the beam storage efficiency. Because
of the variance in the muon momentum, not all muons are at the {\em
  magic momentum} and a small correction must be applied to
$\omega_a$.  The positions of the electrodes are surveyed to the
sub-millimeter level and the electric field generated by the
electrodes are modeled by Opera3D. The electrodes are designed to be
operated up to 25~kV. Dedicated procedures have been developed to
condition the electrodes and to date data has been taken with
potentials between 15 and 21~kV.

\subsection{Detector system}

Twenty-four calorimeter stations are evenly distributed along the
storage ring on the inner side of the vacuum chamber. Decay positrons
exit the vacuum chamber through its aluminum wall and immediately
enter the PbF$_2$ crystals of the calorimeters. The high-energy
positron generates a shower of Cherenkov light in the crystal, and the
number of photons are proportional to the energy of the primary
positron. At the end of each crystal the Cherenkov photons are
detected by a silicon photomultiplier (SiPM), and the signal from the
SiPM is amplified and then read out by a 800-MSPS 12-bit waveform
digitizer. Each calorimeter station has a 6(V)$\times$9(H) array of
crystals. Signals from all crystals are continuously digitized for
700~$\mu$s after the trigger signal synchronized to the incoming muon
bunch. For the T-method, to reduce the data size written to the disk,
only the waveform section near a SiPM signal pulse is saved. The pulse
finding algorithm is implemented using Graphic Processor Units
(GPU). The multiplicities and positions of the particles hitting a
calorimeter station can be determined by the topology of the trigger
pattern and the signal amplitude distribution. Pileup pulses separated
larger than $\sim$4~ns are also resolvable in the E989
calorimeters. The gain stability of each SiPM is monitored using a
dedicated laser-calibration system~\cite{ANASTASI201543}. The response
of the calorimeter system was thoroughly characterised in two SLAC
testbeams. The calorimeter stations together with the
laser calibration system, data acquisition system and slow control
system was finished before May 2017. 

Charged particle straw tracking detectors \cite{TomThesis2018} were
designed to reconstruct the trajectories of positrons before they
enter the calorimeters. One tracker station was constructed and
commissioned in May 2017, and a second tracker station was
commissioned in December 2017. Each tracker station comprises of 8
modules, and each module has two planes of straws oriented at
7.5$^{\circ}$ with respect each other, so that both the horizontal and
vertical coordinates of the trajectory can be determined. The straws
are made of 15-$\mu$m thick aluminized mylar foils with a gold-plated
tungsten wire (d=25\,$\mu$m) at the center. Each straw is filled with
an Argon-Ethane(50\%) gas mixture. A resolution of 165~$\mu$m in plane
perpendicular to the straw has been achieved. Tracks are extrapolated
forward to match with the calorimeter and to the point of tangency of
the ideal muon orbit allowing the profile of the beam distribution to
be determined.

There are also 3 more detector systems used to measure the muon
beam. The T0 detector is a scintillator with two photomultiplier
tubes. It is placed at the end of the delivery beam line before the
muons enters the magnet yoke. It gives the time when the muon bunch
arrives, and this timing information is important for tuning the
kicker timing. The integrated pulse amplitude of the T0 signal is also
a good proxy for the total number of muons in that bunch. The inflector
beam monitors (IBMS) are scintillator fiber arrays in the muon beam
near the inflector. Two IMBS detectors are commissioned. They are
located upstream from the inflector at atmosphere, providing beam
profiles and positions before the muons entering the inflector. Inside
the muon storage ring, there are also two stations of scintillation
fiber detectors for profiling the muon beam. They are useful for
studying the beam dynamics, but destructive for the muon beam. During
production runs, the scintillation fiber detectors in the muon storage
ring are extracted from the beam storage region.

\subsection{Magnetic field measurement system}

The magnetic field in the muon storage region is scanned by a trolley
that moves on the rails inside the vacuum chamber, which carries 17
NMR probes and the electronics for digitizing the NMR waveform. The
trolley is pulled by two cables on each side to move, and the cables
are wound on motorized drums. One of the cables is also the signal
cable for the NMR electronics. The trolley shell was the one used in
E821 \cite{PhysRevD.73.072003}. The motion control is fully automated
and the new NMR electronics allows raw NMR waveforms to be stored and
analysed.  The muon beam is turned off during the magnetic field
scanning. During normal runs with the muon beam, the trolley is
extracted from the storage region but still stays inside the vacuum
chamber. During the stable run period two field scans were performed
per week.

There are 378 NMR probes installed at fixed locations around the ring
to monitor the drift of the magnetic field \cite{Matthias2017}. They
are installed in grooves on the vacuum chambers, above and below the
muon storage region. The readings from these fixed probes are used to
predict the field in the muon storage region, and also used for
the current feedback mechanism to maintain a stable field. The NMR
probes for the trolley and fixed probe systems use petroleum jelly as
the detection material. They were designed and built before 2015, used
in the shimming campaign in 2016 and then installed and commissioned
in 2017. The electronics system for the fixed probes was installed and
commissioned in May 2017, and then several significant upgrades were
implemented from July to December in 2017. The NMR frequency
extraction algorithm is also implemented on GPUs, and the total time
for reading and analyzing waveforms from all fixed probes is 1.67~s,
compared to $\sim$10~s in E821.

The magnetic field read by the NMR probes carried by the trolley is
perturbed by the material of the trolley shell and electronics, and
the material itself is also magnetized in the field. Because the
protons in the NMR probes are in molecules, their NMR frequency is
proportional to, but different from, the free proton NMR frequency. To
correct for these effect, we built a cylindrical probe with pure water
as the detection material. This probe has a well-known magnetic
susceptibility and chemical shift \cite{RevModPhys.88.035009}. The perturbation
of the probe itself was measured. Each NMR probe on the trolley has to
be calibrated to this calibration probe, and dedicated calibration
runs were conducted in vacuum. Spherical calibration probes using
water and $^{3}$He as the detection material have also been
fabricated. These calibration probes will be cross-calibrated to
perform a consistency check.

\section{The Run-1 Physics Data Accumulation}
\label{Sec_PhysRun1}

The commissioning run from May 2017 to July 2017 allowed the
performance of the beamline, injection and detector systems to be
evaluated.  The characteristic signature of $\omega_{a}$ was observed in
the data, but the muon storage efficiency was significantly below the
design. Throughout the summer and with the first beam in November 2017
considerable optimisations and improvements were implemented and the
first physics quality run, Run-1, commenced at the end of March 2018.

\begin{figure}[h!tb]
\centering
\includegraphics[width=0.6\linewidth]{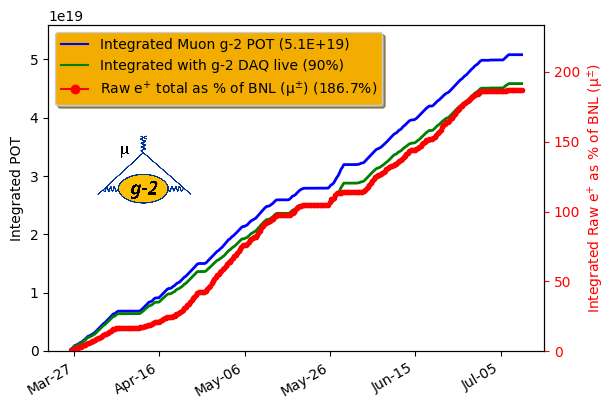}
\caption{Integrated numbers of proton-on-target and the raw number of detected decay positrons recorded during Run-1.}
\label{Fig_Progress}
\end{figure}

A dataset exceeding that accumulated by E821 was collected before the
end of May 2018. During stable running approximately 500 decay
positrons per circulating muon bunch were recorded. The data
acquisition system recorded data at up to 250 Mb/sec and was live for
$\sim$90\% of the beam-on time. The storage ring magnet was
operational over 95\% of the time and more than 30 magnetic field scans were
successfully conducted. The reference clock frequency was blinded for the duration
of the Run-1 period. The full analysis of this Run-1 dataset is expected to be
completed in the second half of 2019.

\begin{figure}[h!tb]
\centering
\includegraphics[width=0.6\linewidth]{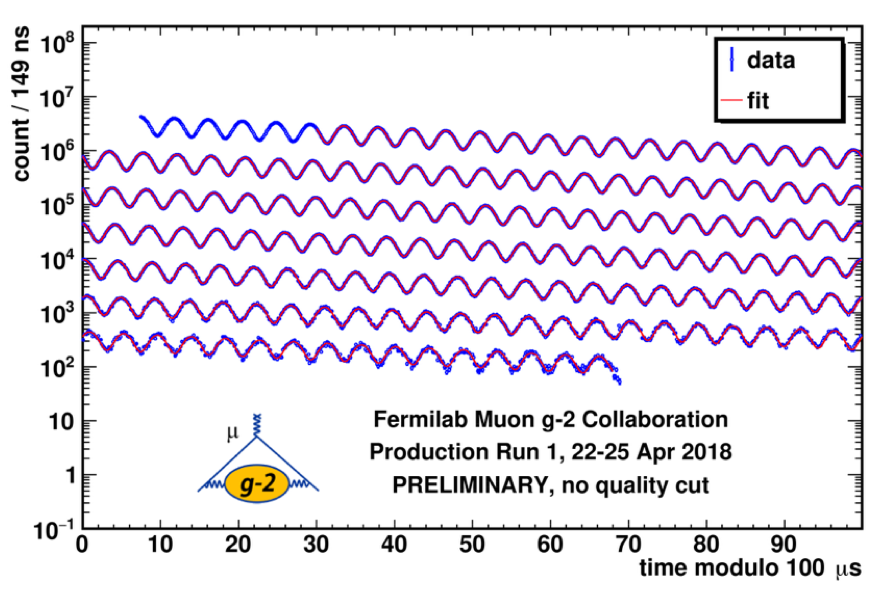}
\caption{(PRELIMINARY) Positron counting rate spectrum fitted to $N_{0} \exp(-t/\tau)(1+A\cos(\omega_{a}t+\phi))$. No quality controls are implemented.}
\label{Fig_Omega_a}
\end{figure}

\begin{figure}[h!tb]
\centering
\includegraphics[width=0.8\linewidth]{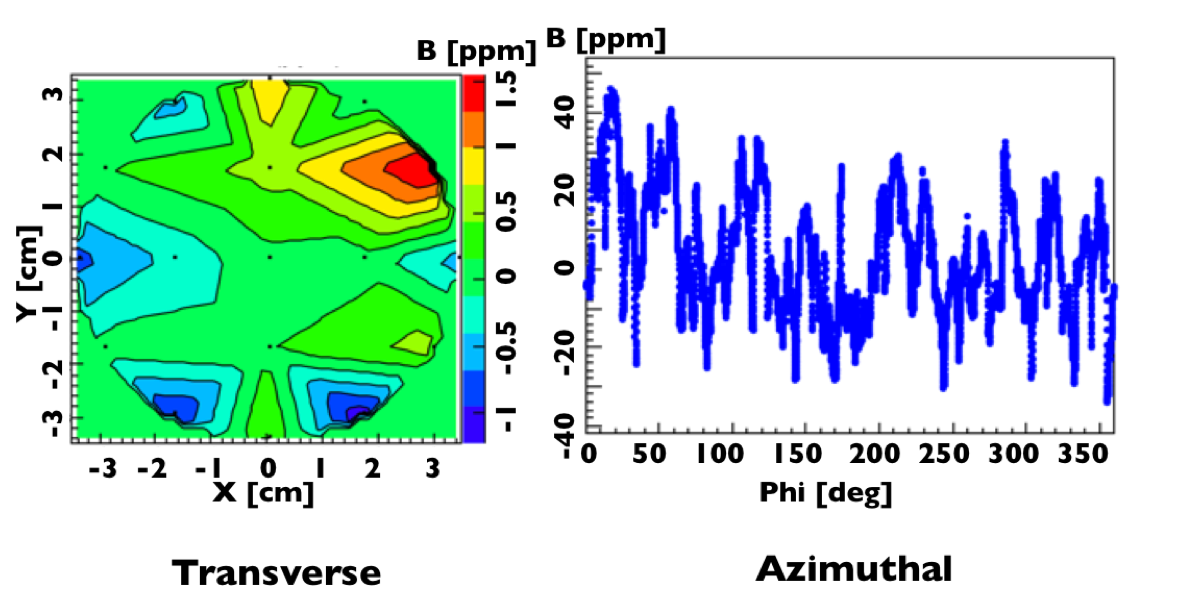}
\caption{(PRELIMINARY) Magnetic field map scanned on May 16th
  2018. The transverse magnetic field distribution is averaged over
  the azimuthal direction, while each data point in the azimuthal
  magnetic field distribution is averaged in the transverse direction.
  No calibration, drift correction and quality controls are
  implemented.}
\label{Fig_Omega_p}
\end{figure}

In Run-1, the number of stored muons was approximately 50\% of the
design and a number of factors have been identified to increase the
number of stored muons towards the design goal and particularly to
improve the momentum variance of the beam. Upgrades to the kicker and
electric quadrupole systems will be undertaken during the summer of
2018 ready for the start of the next physics data taking period in
November 2018. In addition fiber-glass insulation and improved air
conditioning will be installed to ensure that temperature fluctuations
in the experimental hall are reduced which in turn reduces the variance in the
magnetic field and calorimeter gain.

A higher power air conditioning system in the experiment hall will
maintain the temperature better and thus reduce the drift of the
magnetic field and the detector gains. Fiberglass insulation will also
be installed onto the magnet to achieve higher magnetic field
stability.

\section{Conclusion}

The Muon g-2 experiment E989 has been successfully commissioned and a
Run-1 dataset exceeding the statistics of the E821 experiment has been
accumulated.  The goal of this experiment is to search for beyond-SM
physics by measuring the anomalous magnetic moment $a_{\mu}$ of muons
with an uncertainty better than 140~ppb. The Run-1 analysis is
expected to measure $a_{\mu}$ with an uncertainty similar to that of
E821. A dataset several times the size of E821 will be accumulated in
2019 with significant improvements in the muon storage efficiency and
beam quality.

\Acknowledgements
We thank David Flay, Erik Swanson, Hogan Nguyen,
Jason Crnkovic, Chris Stoughton, Jarek Kaspar, David Hertzog, and
Jason Hempstead for their expert knowledge about each system of the
muon g-2 experiment. We thank the Fermilab management and staff for their strong support of this experiment. 

The Muon $g-2$ experiment was performed at the Fermi National
Accelerator Laboratory, a U.S. Department of Energy, Office of
Science, HEP User Facility. Fermilab is managed by Fermi Research
Alliance, LLC (FRA), acting under Contract No. DE-AC02-07CH11359.
Additional support for the experiment was provided by the Department
of Energy offices of HEP and NP (USA), the National Science Foundation
(USA), the Istituto Nazionale di Fisica Nucleare (Italy), the Science
and Technology Facilities Council (UK), the Royal Society (UK), the
European Union's Horizon 2020 research and innovation programme under
the Marie Sk\l{}odowska-Curie grant agreements No. 690835 (MUSE),
No. 734303 (NEWS), MOST and NSFC (China), MSIP and NRF (Republic of
Korea).

\bibliographystyle{unsrt}
\bibliography{g2bibliography_abbrv}



\end{document}